\documentclass[preprint]{ptephy_v1}
\pdfoutput=1
\preprintnumber{KYUSHU-HET-320}
\usepackage{hyperref}
\usepackage{physics}
\usepackage{comment}
\usepackage{tikz}

\begin{document}

\title{'t~Hooft line in 4D $U(1)$ lattice gauge theory and 

a microscopic description of dyon's statistics}

\author{Soma Onoda}
\affil{\email{onoda.soma@phys.kyushu-u.ac.jp}}
\affil{Department of Physics, Kyushu University, 744 Motooka, Nishi-ku,
Fukuoka 819-0395, Japan}

\begin{abstract}%
In lattice gauge theory with compact gauge field variables, an introduction of
the gauge field topology requires the assumption that lattice field
configurations are sufficiently smooth. This assumption is referred to as the
admissibility condition. However, the admissibility condition always ensures
the Bianchi identity, and thus prohibits the existence of magnetic objects such
as the 't~Hooft line. Recently, in 2D compact scalar field theory,
Ref.~\cite{Abe:2023uan} proposed a method to define magnetic objects without
violating the admissibility condition by introducing holes into the lattice. In
this paper, we extend this ``excision method'' to 4D Maxwell theory and
propose a new definition of the 't~Hooft line on the lattice. Using this
definition, we first demonstrate a lattice counterpart of the Witten effect
which endows the 't~Hooft line with electric charge and make it a dyon.
Furthermore, we show that by interpreting the 't~Hooft line as a boundary of
the lattice system, the statistics of the dyon can be directly read off. We
also explain how the dyonic operator which satisfies the Dirac quantization
condition becomes a genuine loop operator even at finite lattice spacings.
\end{abstract}

\subjectindex{B01, B04, B27, B31}

\maketitle

\section{Introduction}
Lattice regularization provides a non-perturbative definition of quantum field
theories while preserving gauge symmetry exactly. It is not only a practical
tool for numerical simulations, but also plays an essential role in
formulating various concepts in quantum field theory in a non-perturbative
manner. In this paper, we discuss the 't~Hooft line~\cite{tHooft:1977nqb} and
the associated topological aspects of quantum field theory.

The 't~Hooft line is the world line of a monopole. It serves as an order parameter that distinguishes the confined phase of monopoles, and is also related to
quark confinement via the dual superconductivity
picture~\cite{tHooft:1981bkw,Seiberg:1994rs}. Furthermore, in the presence of a
topological term known as the $\theta$~term (or axion-photon coupling), it is
known that a shift in the $\theta$~angle causes the monopole to acquire
electric charge and become a dyon---a phenomenon referred to as the Witten
effect~\cite{Witten:1979ey}. Interestingly, the statistics of the resulting
dyon is known to deviate by~$+1$ from the naive sum of the statistics of the
constituent electric and magnetic charges; this caused by the contribution of
the angular momentum of the electromagnetic
field~\cite{Goldhaber:1976dp,Jackiw:1976xx,Jackiw:1981wc}. For example, even if
both the electrically charged and magnetically charged particles are bosons,
their bound state---the dyon---can behave as a fermion. This phenomenon also
occurs in the context of the Witten effect; when a bosonic 't~Hooft line is
present, a shift of the $\theta$ angle leads to a dyonic line that acquires
nontrivial statistics~\cite{metlitski2013bosonic}. In this paper, we
demonstrate within the framework of lattice gauge theory that the 't~Hooft line
acquires nontrivial statistics through the Witten effect (See also related
recent works~\cite{Sulejmanpasic:2019ytl,Jacobson:2023cmr,Aoki:2023lqp,%
Honda:2024sdz,Honda:2024xmk,Kobayashi:2024dqj}).

The $\theta$ term is intrinsically related to the topology of gauge fields. In
lattice gauge theories formulated with compact field variables, the
incorporation of topological properties typically requires the assumption that
lattice gauge field configurations are sufficiently
smooth~\cite{Luscher:1981zq}. This assumption is known as the
\emph{admissibility condition}. Under this condition, field configurations can
be classified into topological sectors~\cite{Luscher:1981zq,Luscher:1998du}.
More precisely, admissibility allows one to define principal bundles purely
from lattice degrees of freedom~\cite{Luscher:1981zq}, and it also ensures the
locality of the overlap Dirac
operator~\cite{Neuberger:1998wv,Hernandez:1998et}. Furthermore, since the
admissibility condition enforces the Bianchi identity at all lattice sites, it
effectively prohibits (dynamical) monopoles in theories such as $PSU(N)$
Yang--Mills and $U(1)$ Maxwell by rendering them infinitely heavy.

In this work, we consider the 4D $U(1)$ Maxwell theory on the lattice with
compact variables. In this setup, the Bianchi identity implied by
admissibility leads to a magnetic 1-form symmetry. However, the introduction of
monopoles---the charged objects under this symmetry---as probes, generically
conflicts with the admissibility condition. In other words, the requirement of
smoothness excludes singularities such as magnetic objects.

A method to introduce magnetic objects was proposed in the context of 2D
compact scalar field theory in~Ref.~\cite{Abe:2023uan}. In that work, so-called
magnetic objects, where the scalar field carries a local winding number, were
defined by introducing holes into the lattice---an approach referred to as the
\emph{excision method}. Furthermore, by introducing the dual lattice, the
action of the winding symmetry on magnetic objects, as well as the Witten
effect, was derived.

In this work, we extend the excision method developed
in~Ref.~\cite{Abe:2023uan} for 2D compact scalar field theory to the 4D Maxwell
theory. However, in the case of 4D Maxwell theory, it turns out that
introducing a dual lattice with appropriate degrees of freedom is technically
challenging. As a result, the Witten effect cannot be derived in a fully
rigorous manner, unlike the case of 2D scalar field theory
in~Ref.~\cite{Abe:2023uan}.

In this work, we define the 't~Hooft line by introducing a hole into the
lattice according to the excision method. In other words, we introduce a
boundary into the lattice system, whose topology is given
by~$\partial M_4\cong S^1\times S^2$. In the presence of the $\theta$~term,
$i\theta/(8\pi^2)\sum_{M_4}f\cup f$, a Chern--Simons term emerges on the
boundary. This Chern--Simons term can be interpreted as a ``bundle'' of Wilson
lines whose total number equals the magnetic charge. We demonstrate that
reflecting the fermionic nature of the Chern--Simons
theory, these Wilson lines exhibit nontrivial statistics.

In~Sect.~\ref{set-up}, we describe the definition of fields on a lattice and
introduce the definition of the 't~Hooft line based on the excision method.
In~Sect.~\ref{Witten}, we incorporate the $\theta$~term and demonstrate that
the Witten effect can be reproduced in the continuum limit using the definition
of the 't~Hooft line introduced in~Sect.~\ref{set-up}. On the other hand,
Sect.~\ref{stat} investigates the nontrivial statistics of dyons without taking
the continuum limit, by examining the gauge invariance of the surface term in
the $\theta$~term. We further show that when the electric charge induced by the
Witten effect is an integer, one obtains a genuine dyonic line, which exhibits
fermionic statistics. Section~\ref{conclusion} provides a summary and
discussion. In Appendix~\ref{integer}, we prove that the so-called topological
charge takes integer values even on the lattice, using the viewpoint of the
Poincar\'e dual of the magnetic flux.

\section{Formulation of compact $U(1)$ lattice gauge theory}
\label{set-up}
\subsection{Set up}
Let us consider $U(1)$ Maxwell theory on $T^4$ approximated by a square
lattice. We denote this four-dimensional lattice system by~$M_4$. The degrees
of freedom are defined on various lattice elements: hypercubes~($h$),
cubes~($c$), plaquettes~($p$), links~($\ell$), and sites~($n$). The fundamental
dynamical variable is the compact link variable, defined by
\begin{align}
   u_\mu(n)=e^{ia_\mu(n)},\qquad-\pi\leq a_\mu(n)<\pi,\qquad\mu=1,2,3,4.
\label{eq:def_link}
\end{align}
Here, $n$ denotes a lattice site and $\mu$ denotes the direction. The gauge
potential $a_\mu(n)$ is defined as the logarithm of the link
variable~$u_{\mu}(n)$.

The field strength of the $U(1)$ gauge field is defined by taking the logarithm
of the plaquette term:
\begin{align}
   f_{\mu\nu}(n)&\equiv\frac{1}{i}\ln u_p
\notag\\
   &=\frac{1}{i}\ln\left[
   u_\mu(n)u_\nu(n+\hat{\mu})u_\mu(n+\hat{\nu})^*u_\nu(n)^*\right]
\notag\\
   &=\Delta_\mu a_\nu(n)-\Delta_\nu a_\mu(n)+2\pi z_{\mu\nu}(n),
   \qquad-\pi\leq f_{\mu\nu}(n)<\pi.
\label{field strength}
\end{align}
Here, $\Delta_\mu$ denotes the forward difference, defined
by~$\Delta_\mu a_\nu(n)=a_\nu(n+\hat{\mu})-a_\nu(n)$.\footnote{%
Throughout this paper, $\hat{\mu}$ denotes the unit vector in
$\mu$~direction.} The integer $z_{\mu\nu}(n)$ ensures that $f_{\mu\nu}(n)$ is
within the principal branch~$[-\pi,\pi)$.

Under the $U(1)$ gauge transformation, a link variable transforms as
\begin{align}
   u_\mu(n)\to e^{-i\lambda(n)}\,u_\mu(n)\,e^{i\lambda(n+\hat{\mu})}.
\end{align}
Correspondingly,
\begin{align}
   a_\mu(n)&\to a_\mu(n)+\Delta_\mu\lambda(n)+2\pi m_\mu(n),
\\
   z_{\mu\nu}(n)&\to z_{\mu\nu}(n)
   -\left\{\Delta_\mu m_\nu(n)-\Delta_\nu m_\mu(n)\right\},
\end{align}
where $m_\mu(n)$ is an integer chosen to
ensure~$-\pi\leq a_\mu(n)+\Delta_\mu\lambda(n)+2\pi m_\mu(n)<\pi$.

We define the operator that measures the magnetic flux as follows:
\begin{align}
   Q_{\text{mag}}(S)
   &\equiv\frac{1}{2\pi}\sum_{(n,\mu\nu)\in S}f_{\mu\nu}(n)
\notag\\
   &=\sum_{(n,\mu\nu)\in S}z_{\mu\nu}(n)\in\mathbb{Z},
\label{eq:def_Qmag}
\end{align}
where $S$ is an oriented closed 2-surface, and by definition, $Q_{\text{mag}}(S)$
takes an integer value.

In this paper, to endow the lattice theory with topological natures, we impose
a sufficient smoothness condition on the lattice gauge fields, known as the
\emph{admissibility condition}~\cite{Luscher:1981zq,Hernandez:1998et}:
\begin{align}
   \sup_{n,\mu,\nu}\left|f_{\mu\nu}(n)\right|<\epsilon,
   \qquad0<\epsilon<\frac{\pi}{3}.
\label{admissibility}
\end{align}
Under this condition, we have
\begin{align}
   \frac{1}{2\pi}\left|(\dd f)_c\right|
   &=\frac{1}{2\pi}
   \left|\frac{1}{2}\sum_{\nu,\rho,\sigma}\epsilon_{\mu\nu\rho\sigma}
   \Delta_\nu f_{\rho\sigma}(n)\right|
\notag\\
   &=\left|\frac{1}{2}\sum_{\nu,\rho,\sigma}\epsilon_{\mu\nu\rho\sigma}
   \Delta_\nu z_{\rho\sigma}(n)\right| 
   <\frac{3\epsilon}{\pi}<1.
\end{align}
Here, $\dd$ denotes the coboundary operator (see~Ref.~\cite{Jacobson:2023cmr}
for a precise definition), and $c$ refers to the cube extending in directions
complementary to~$\mu$ from a site~$n$. Since $z_{\mu\nu}(n)$ is an integer and
the left-hand side is strictly less than 1, it must be~$0$. Thus, the lattice
Bianchi identity $(\dd f)_c=0$ holds.

Under this identity, $Q_{\text{mag}}(S)$ is shown to be topological because
\begin{align}
   Q_{\text{mag}}(S)-Q_{\text{mag}}(S') 
   =\frac{1}{2\pi}\sum_{p\in S}f_p-\frac{1}{2\pi}\sum_{p\in S'}f_p 
   &=\frac{1}{2\pi}\sum_{p\in S\cup\bar{S'}}f_p 
   =\frac{1}{2\pi}\sum_{c\in D}(\dd f)_c=0,
\label{topological deform}
\end{align}
where $f_p$ denotes $f_{\mu\nu}(n)$ defined on a plaquette~$p$ and the summation
domain~$D$ is such that $\partial D=S\cup\bar{S'}$.
Eq.~\eqref{topological deform} shows that the admissibility condition
guarantees the magnetic 1-form symmetry in the $U(1)$ Maxwell theory.

\subsection{'t~Hooft line on the lattice with admissibility condition}\label{excision-method}
Having constructed the generator of the magnetic 1-form symmetry,
$Q_{\text{mag}}(S)$, we now aim to define the corresponding charged object---the
’t~Hooft line (the worldline of a magnetic monopole)---in the lattice
formulation.

Since gauge field configurations that satisfy the admissibility condition
always obey the Bianchi identity, one may say that magnetic objects such as
’t~Hooft lines, which would correspond to violations of the Bianchi identity,
are forbidden in this setup. To address this issue,
Refs.~\cite{Abe:2023uan,Morikawa:2024zyd} proposed the definition of magnetic
objects in compact scalar field theories by excising a part of the lattice
while preserving the admissibility condition. In this paper, we extend this
\emph{excision method} to 4D $U(1)$ Maxwell theory to define the ’t~Hooft line.

To introduce a monopole, we select a region $\mathcal{B}$ within a
three-dimensional hypersurface and remove all lattice sites contained
in~$\mathcal{B}$. We regard this region $\mathcal{B}$ as a monopole, and define
its magnetic charge $m$ as:
\begin{align}
   m\equiv Q_{\text{mag}}(\partial\mathcal{B})
   =\frac{1}{2\pi}\sum_{p\in\partial\mathcal{B}}f_p
   =\sum_{p\in\partial\mathcal{B}}z_p.
\label{def:magnetic-charge}
\end{align}
Here, due to the admissibility condition~\eqref{admissibility}, the region
$\mathcal{B}$ must be sufficiently large as follows:
\begin{align}
   (\text{the number of plaquettes on $\partial\mathcal{B}$})
   >\frac{2\pi|m|}{\epsilon}
   >\frac{2\pi}{\epsilon}.
\label{sizebound}
\end{align}
However, since the bound~\eqref{sizebound} does not depend on the lattice
spacing, $\mathcal{B}$ can be point-like in the continuum limit, and thus can
be regarded as a monopole in the continuum theory.

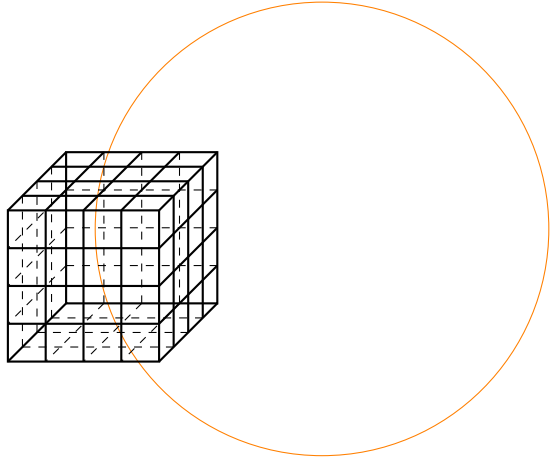
\begin{figure}
\centering
\begin{tikzpicture}[scale=2]
\begin{scope}
\draw[thin,orange] (1.5,0.3,-0.5) circle[radius=1.5cm];
     
        \draw[thick] (0,0,0) -- (1,0,0) -- (1,1,0) -- (0,1,0) -- cycle; 
        \draw[thick] (0,0,1) -- (1,0,1) -- (1,1,1) -- (0,1,1) -- cycle; 
        \draw[thick] (0,0,0) -- (0,0,1);
        \draw[thick] (1,0,0) -- (1,0,1);
        \draw[thick] (1,1,0) -- (1,1,1);
        \draw[thick] (0,1,0) -- (0,1,1);

        \foreach \x in {0.25, 0.5, 0.75} {
            \draw[dashed] (\x,0,0) -- (\x,1,0); 
            \draw[thick] (\x,0,1) -- (\x,1,1);
            \draw[dashed] (\x,0,0) -- (\x,0,1); 
            \draw[thick] (\x,1,0) -- (\x,1,1);
        }
        \foreach \y in {0.25, 0.5, 0.75} {
            \draw[dashed] (0,\y,0) -- (1,\y,0); 
            \draw[thick] (0,\y,1) -- (1,\y,1); 
            \draw[dashed] (0,\y,0) -- (0,\y,1);
            \draw[thick] (1,\y,0) -- (1,\y,1);
        }
        \foreach \z in {0.25, 0.5, 0.75} {
            \draw[dashed] (0,0,\z) -- (1,0,\z); 
            \draw[thick] (0,1,\z) -- (1,1,\z); 
            \draw[dashed] (0,0,\z) -- (0,1,\z); 
            \draw[thick] (1,0,\z) -- (1,1,\z); 
        }
    \end{scope}
\end{tikzpicture}
\caption{To define a monopole, a region is excised from a certain time-slice
of the lattice configuration as shown in this figure. The orange circle
represents a loop extending in the direction orthogonal to this time-slice.}
\label{fig:tHooftloop}
\end{figure}

To obtain a ’t~Hooft line carrying magnetic charge~$m$, we prepare a
worldline~$\ell$, and place a region~$\mathcal{B}$ on each three-dimensional
hypersurface orthogonal to~$\ell$~(see Fig.~\eqref{fig:tHooftloop}). Thus, the ’t~Hooft line can be regarded as
a boundary with the lattice topology $S^2\times\ell$.\footnote{We assume
that $\partial\mathcal{B}\cong S^2$.} In the case of a ’t~Hooft loop, we take
$\ell\cong S^1$. To define a monopole with a specific magnetic charge, one must
impose boundary conditions on~$\partial\mathcal{B}$ such that
Eq.~\eqref{def:magnetic-charge} holds. Once Eq.~\eqref{def:magnetic-charge} is
satisfied at a specific ``time slice'', the Bianchi identity ensures that the
magnetic charge is conserved at all times.

\section{4D Maxwell theory with $\theta$ term and the Witten effect}
\label{Witten}
Using the field strength defined in~Eq.~\eqref{field strength}, we define the
lattice action of the 4D Maxwell theory as follows:
\begin{align}
   S_{\text{Maxwell}}
   \equiv\sum_n\sum_{\mu,\nu}
   \frac{1}{4g_0^2}f_{\mu\nu}(n)f_{\mu\nu}(n)-S_\theta.
\end{align}
Here, $g_0$ denotes the bare gauge coupling, and $S_\theta$ is the topological
term with the $\theta$ angle~\cite{Fujiwara:2000wn}:
\begin{align}
   S_\theta\equiv\frac{i\theta}{32\pi^2}\sum_n\sum_{\mu,\nu,\rho,\sigma}
   \epsilon_{\mu\nu\rho\sigma}f_{\mu\nu}(n)f_{\rho\sigma}(n+\hat{\mu}+\hat{\nu})
   =\frac{i\theta}{8\pi^2}\sum_{h\in M_4}(f\cup f)_h.
\label{theta term}
\end{align}
Here, $h$ denotes a hypercube of the four-dimensional lattice~$M_4$. The cup
product~$\cup$ is defined on the hypercubic lattice as
in~Refs.~\cite{Chen:2021ppt,Jacobson:2023cmr}.\footnote{The cup product
satisfies a Leibniz rule analogous to the wedge product in continuum theory:
\begin{align}
   \dd(\alpha^{(p)}\cup\beta^{(q)})
   =\dd\alpha^{(p)}\cup\beta^{(q)}+(-1)^p\alpha^{(p)}\cup\dd\beta^{(q)},
\end{align}
where $\alpha^{(p)}$ and $\beta^{(q)}$ denote $p$- and $q$-cochains. However, the
cup product is not commutative at the cochain level:
\begin{align}
   \alpha^{(p)}\cup\beta^{(q)}-(-1)^{pq}\beta^{(q)}\cup\alpha^{(p)}
   =(-1)^{p+q+1}\left\{\dd(\alpha^{(p)}\cup_1\beta^{(q)})
   -\dd\alpha^{(p)}\cup_1\beta^{(q)}-(-1)^p\alpha^{(p)}\cup_1\dd\beta^{(q)}
   \right\},
\label{cup-product-noncommutative}
\end{align}
where $\cup_1$ denotes the cup-1 product (a higher cup
product)~\cite{Chen:2021ppt,Jacobson:2023cmr}.}

When the four-dimensional Euclidean lattice system~$M_4$ has no boundary, the
topological term $S_\theta$ becomes
\begin{align}
   S_\theta
   &=\frac{i\theta}{8\pi^2}\sum_{h\in M_4}(f\cup f)_h
\notag\\
   &=\frac{i\theta}{8\pi^2}\sum_{h\in M_4}
   \left\{\dd(a\cup\dd a+2\pi a\cup z+2\pi z\cup a)\right\}_h
   +\frac{i\theta}{2}\sum_{h\in M_4}(z\cup z)_h
\notag\\
   &=\frac{i\theta}{2}\sum_{h\in M_4}(z\cup z)_h.
\end{align}
Since $M_4$ has no boundary, the total derivative term in the second line
vanishes. When $M_4=T^4$, $(i/2)\sum_{h\in M_4}(z\cup z)_h$ is an integer
(see~Appendix~\ref{integer}), as the continuum counterpart for spin manifolds.
Therefore, the $\theta$~angle in~Eq.~\eqref{theta term} has $2\pi$-periodicity.

Now, when an 't~Hooft loop carrying magnetic charge~$m$ is present, the
spacetime~$M_4$ has a boundary~$\partial M_4\cong S^2\times S^1$. In this case,
the topological term~$S_\theta$ becomes
\begin{align}
   S_\theta
   =\frac{i\theta}{4\pi}\sum_{c\in\partial M_4}
   \left(\frac{a\cup\dd a}{2\pi}+a\cup z+z\cup a\right)_c
   +\frac{i\theta}{2}\sum_{h\in M_4}(z\cup z)_h.
   \label{CS}
\end{align}
The boundary term can be interpreted as a $U(1)$ Chern--Simons term. From the
Bianchi identity~$(\dd f)_c=(\dd z)_c=0$, the Poincar\'e dual of the
2-cocycle~$z$ on~$\partial M_4$ must be a closed 1-cycle~$\ell_1$
as\footnote{$\delta(\ell_1)$ is defined such that
$\sum_{c\in\partial M_4}(a\cup\delta(\ell_1))_c=\sum_{\ell\in\ell_1}a$ holds.}
\begin{align}
   z|_{\partial M_4}=\delta(\ell_1),\qquad
   \dd z|_{\partial M_4}=\delta(\partial\ell_1)=0.
\label{eq:poincare-dual-3d}
\end{align}
Using this, we find
\begin{align}
   \frac{i\theta}{4\pi}
   \sum_{c\in\partial M_4}(a\cup z+z\cup a)_c
   =\frac{i\theta}{4\pi}\sum_{c\in\partial M_4}
   (a\cup\delta(\ell_1)+\delta(\ell_1)\cup a)_c.
\label{eq:framed-Wilson-lines}
\end{align}
Thus, the term $i\theta/(4\pi)\sum_{c\in\partial M_4}(a\cup z+z\cup a)_c$
in~Eq.~\eqref{CS} can be interpreted as a collection of framed Wilson
lines~\cite{Jacobson:2023cmr} along~$\ell_1$ on the boundary~$\partial M_4$.
This set of lines includes both non-contractible 1-cycles~$\ell_1^{\text{non}}$
and contractible ones~$\partial S_2$, namely,
\begin{align}
   z|_{\partial M_4}=\delta(\ell_1)
   =\delta(\ell_1^{\text{non}})+\delta(\partial S_2)
   =\delta(\ell_1^{\text{non}})+\dd\delta(S_2).
\end{align}
The contribution from the contractible 1-cycle can be absorbed by a gauge
transformation $z\to z-\dd r$ and hence is unphysical. On the other hand, the
contribution from the non-contractible 1-cycle cannot be removed by a gauge
transformation. This represents the magnetic flux carried by the 't~Hooft loop.
In other words, the net number of non-contractible 1-cycles is equal to the
magnetic charge~$m$ and they wrap around the $S^1$ of the 't~Hooft loop as
\begin{align}
   \delta(\ell_1^{\text{non}})=\sum_iq_i\delta(\ell_1^{\text{non},i}),
   \qquad\sum_iq_i=m,
\end{align}
where $q_i\in\{-1,1\}$, and each $\ell_1^{\text{non},i}\cong S^1$ corresponds to a
part of the Poincar\'e dual of~$z$ contributing to magnetic charge by~$+1$.
Actually, the expression~\eqref{eq:framed-Wilson-lines} implies the Witten
effect. To see this, we note that the configuration of~$z$
in~Eq.~\eqref{eq:poincare-dual-3d} can be extended into the bulk~$M_4$ as
\begin{align}
   z=\sum_iq_i\delta(\mathcal{R}_i)+z_0,\qquad\sum_iq_i=m,
\label{z-separate}
\end{align}
where each $\mathcal{R}_i$ is a 2-disk and it satisfies
$\ell_1^{\text{non},i}=\partial\mathcal{R}_i$; $z_0$ is given by the delta
function whose support is a union of closed surfaces on~$M_4$.\footnote{%
$\delta(\mathcal{R})$ is defined such that
$\sum_{h\in M_4}(f\cup\delta(\mathcal{R}))_h=\sum_{p\in\mathcal{R}}f$ for general
2-form~$f$}\footnote{In general, $z$ may include additional components
supported on 2-disks not wrapping the $S^1$ of the 't~Hooft loop, but these are
pure-gauge and hence unphysical.}

Now, the topological term becomes
\begin{align}
   S_\theta 
   &=\frac{i\theta}{4\pi}\sum_{c\in\partial M_4}
   \left(\frac{a\cup\dd a}{2\pi}+a\cup z+z\cup a\right)_c 
   +\frac{i\theta}{2}\sum_{h\in M_4}(z\cup z)_h
\notag\\
   &=\frac{i\theta}{4\pi}\sum_{c\in\partial M_4}\frac{(a\cup\dd a)_c}{2\pi} 
   +\frac{i\theta}{4\pi}\sum_{c\in\partial M_4}\sum_iq_i
   \left\{a\cup\delta(\partial\mathcal{R}_i)+\delta(\partial\mathcal{R}_i)
   \cup a\right\}_c
\notag\\
   &\qquad{}
   +\frac{i\theta}{4\pi}\sum_{h\in M_4}
   \sum_iq_i\left\{2\pi z\cup\delta(\mathcal{R}_i)+2\pi\delta(\mathcal{R}_i)
   \cup z\right\}_h
\notag\\
   &\qquad{}
   +\frac{i\theta}{2}\sum_{h\in M_4}(z_0\cup z_0)_h 
   -i\theta\sum_{h\in M_4}\sum_{i,j}q_iq_j\delta(\mathcal{R}_i)
   \cup\delta(\mathcal{R}_j).
\label{CS-w/-bulk-rewrite-R}
\end{align}

First, without loss of generality, we can assume via a gauge transformation
that the Poincar\'e dual of~$z_0$ is sufficiently far from the
boundary~$\partial M_4$. Under this assumption, we can argue, based on the
discussion in~Appendix~\ref{integer}, that the
term~$(1/2)\sum_{h\in M_4}(z_0\cup z_0)_h$ is an integer. Hence, the last line
of~Eq.~\eqref{CS-w/-bulk-rewrite-R} can be interpreted
as~$i\theta\times(\text{integer})$.

Next, the second and third terms on the right-hand side are composed of the
following quantities:
\begin{align}
   w_i&\equiv\frac{i\theta}{4\pi}\sum_{c\in\partial M_4}
   \{a\cup\delta(\partial\mathcal{R}_i)\}_c
   +\frac{i\theta}{4\pi}\sum_{h\in M_4}
   \{2\pi z\cup\delta(\mathcal{R}_i)\}_h
\notag\\
   &=\frac{i\theta}{4\pi}
   \sum_{h\in M_4}\{f\cup\delta(\mathcal{R}_i)\}_h,
\label{unnit-Wilson-line} \\
   w_i'&\equiv\frac{i\theta}{4\pi}\sum_{c\in\partial M_4}
   \{\delta(\partial\mathcal{R}_i)\cup a\}_c
   +\frac{i\theta}{4\pi}\sum_{h\in M_4}\{2\pi\delta(\mathcal{R}_i)\cup z\}_h
\notag\\
   &=\frac{i\theta}{4\pi}\sum_{h\in M_4}
   \{\delta(\mathcal{R}_i)\cup f\}_h,
\label{unnit-Wilson-line2}
\end{align}
where we have used the definition of~$f_p$ as $f_p=(\dd a)_p+2\pi z_p$.
The exponential of these operators, $e^w$ and~$e^{w'}$, can be regarded
as~\emph{non-genuine} Wilson loop operators carrying electric
charge~$\theta/(4\pi)$.

To see the expected form of the Witten effect in the continuum theory, let us
consider the classical continuum limit. Denoting the lattice spacing by~$s$,
let $S_{\text{sphere}}$ be the area of the $S^2$ associated with the 't~Hooft loop
in lattice units and $L_{\text{circle}}$ be the length of the $S^1$:
\begin{align}
   S_{\text{sphere}}=\order{s^0},\qquad
   L_{\text{circle}}=\order{s^{-1}}.
\label{order-loop}
\end{align}
Meanwhile, the magnitude of~$w_i$ scales as
\begin{align}
   w_i\simeq\order{L_{\text{circle}}^2},
   \qquad\text{$w_i-w_j\simeq\order{L_{\text{circle}}}$ for~$i\neq j$}.
\end{align}
Here, we have used the fact, by using the Bianchi identity, $w_i-w_j$ is given
by the sum of~$f_p$ over the side of a cylindrical region bounded
by~$\mathcal{R}_i$ and $\mathcal{R}_j$ as its bases, whose area scales
as~$\order{L_{\text{circle}}}$. Since the
relation~$\order{L_{\text{circle}}}\ll\order{L_{\text{circle}}^2}$ can be expected in
the continuum limit, we obtain\footnote{%
Of course, there may exist extreme field configurations in which the
contribution from the side of the cylinder dominates over those from the bases
$R_i$ and $R_j$. In such a case, the present argument does not hold. However,
in the continuum limit, the degrees of freedom on the disks $R_i$ and~$R_j$
become infinitely larger than those on the side, so such extreme configurations
are expected to be negligible.}
\begin{align}
    w_i\simeq w_j\quad\text{for all $i$ and~$j$}.
\label{continuum-limit-of-w_i}
\end{align}
Furthermore, assuming that $a\cup\dd a/s^3$ remains finite in the continuum
limit~$s\to0$, we obtain
\begin{align}
   \sum_{c\in\partial M_4}\frac{a\cup\dd a}{2\pi}
   \simeq\order{s^3}\times\order{L_{\text{circle}}}\times
   \order{S_{\text{sphere}}}\sim\order{s^2}\to 0,\quad s\to0,
\end{align}
and hence, the first term on the right-hand side of~Eq.~\eqref{CS} becomes
negligible. As a result, we obtain
\begin{align}
   S_\theta\to\frac{im\theta}{2\pi}\int_\mathcal{R}f+i\theta\mathbb{Z},
\label{continuum-wilson-loop}
\end{align}
in the continuum limit, successfully reproducing the Witten effect expected
for finite $\theta$ in the continuum theory. In other words, if we write the
electric and magnetic charges of a line operator as~$(q,m)$, then the
phenomena~$(q,m)\xrightarrow{\theta\to\theta+\Delta\theta}
(q+m\Delta\theta/(2\pi),m)$ is derived from the lattice theory and its
classical continuum limit.

Now, in the expression~\eqref{continuum-wilson-loop} obtained in the continuum
limit, when $m\theta=2\pi\mathbb{Z}$, the term $i\int_\mathcal{R} f$ reduces
to~$i\int_{\partial \mathcal{R}}a\bmod2\pi i$, and thus becomes a \emph{genuine}
loop operator. However, the relation~\eqref{continuum-limit-of-w_i}
and~Eq.~\eqref{continuum-wilson-loop} hold only in the continuum limit.
It is nontrivial therefore whether such a genuine loop operator arises at
finite lattice spacings. In the next section, we demonstrate that even at
finite lattice spacings, a genuine loop operator emerges when
$m\theta=2\pi\mathbb{Z}$.

\section{Microscopic description of dyon's statistics}
\label{stat}
\begin{figure}[htbp]
\centering
\begin{tikzpicture}[scale=2]
\begin{scope}
    \draw[->, thick, red] (-4,0,0) -- (-4,2,0);
    \draw[thick] (-4.5,0,-0.5) -- (-4,0,-0.5) -- (-4,0,-1) -- (-4.5,0,-1) -- cycle; 
    \draw[thick] (-4.5,0.5,-0.5) -- (-4,0.5,-0.5) -- (-4,0.5,-1) -- (-4.5,0.5,-1) -- cycle;
    \draw[thick] (-4.5,1,-0.5) -- (-4,1,-0.5) -- (-4,1,-1) -- (-4.5,1,-1) -- cycle;

    \draw [->,very thick] (-3,1,0) -- (-1,1,0);
    \node[below] at (-2,1,0) {gauge transf. $z \to z - \dd r$};

        \draw[->, thick, red] (0,0,0) -- (0,1,0);
        \draw[->, thick, red] (0,1,0) -- (1.5,1,0);
        \draw[->, thick, red] (1.5,1,0) -- (1.5,1,-1.5);
        \draw[->, thick, red] (1.5,1,-1.5) -- (0,1,-1.5);
        \draw[->, thick, red] (0,1,-1.5) -- (0,1,-0.13);
        \draw[->, thick, red] (0,1,-0.13) -- (0,2,-0.13);

        \draw[->, very thick, blue] (0,0.5,-1) -- (0,1,-1);
        \draw[->, thick, blue] (0.5,0.5,-1) -- (0.5,1,-1);
        \draw[->, thick, blue] (1,0.5,-1) -- (1,1,-1);

        \draw[->, thick, blue] (0,0.5,-1.5) -- (0,1,-1.5);
        \draw[->, thick, blue] (0.5,0.5,-1.5) -- (0.5,1,-1.5);
        \draw[->, thick, blue] (1,0.5,-1.5) -- (1,1,-1.5);

        \draw[very thick] (-0.5,0.5,-0.5) -- (0,0.5,-0.5) -- (0,0.5,-1) -- (-0.5,0.5,-1) -- cycle;

\end{scope}
\end{tikzpicture}
\caption{Gauge transformation of the Poincar\'e dual of~$z$ on~$\partial M_4$
for the case $-z\cup r=1$. The red arrows represent the Poincar\'e dual of~$z$
defined on the dual lattice. The black plaquettes in the left-hand side
correspond to the plaquettes~$p$ with~$z_p=1$, oriented in the direction of the
red arrows. The right-hand side shows the configuration after a gauge
transformation has been applied to the left hand side. The blue arrows indicate
the links~$\ell$, where~$-r_\ell = 1$. Since the thick black plaquette and the
thick blue link intersect once~$-z\cup r=1$.}
\label{fig:twist}
\end{figure}
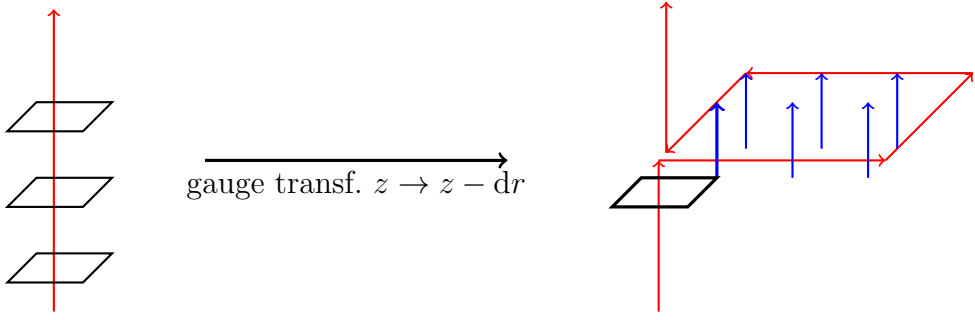

In this section, we discuss the Witten effect without taking the classical
continuum limit. We consider a $U(1)$ gauge
transformations~$a\to a+\dd\lambda+2\pi r$ and~$z\to z-\dd r$ on the boundary
terms in~$S_\theta$, Eq.~\eqref{eq:framed-Wilson-lines}. Under these
transformations, the boundary term transforms as
\begin{align}
   &\frac{i\theta}{4\pi}\sum_{c\in\partial M_4}
   \left(\frac{a\cup\dd a}{2\pi}+a\cup z+z\cup a\right)_c
\notag\\
   &\qquad{}
   \to\frac{i\theta}{2}\sum_{c\in\partial M_4}
   \left(\frac{a\cup \dd a}{2\pi}+a\cup z+z\cup a\right)_c 
   +\frac{i\theta}{2}\sum_{c\in\partial M_4}
   \left(r\cup z+z\cup r-r\cup\dd r\right)_c.
\label{theta-stat}
\end{align}
The gauge transformation effectively adds a contractible loop to the
Poincar\'e dual of~$z$. In other words, if $-r|_{\partial M_4}=\delta(S_2)$,
then the transformation on~$z$ becomes
$z|_{\partial M_4}=\delta(\ell_1)\to\delta(\ell_1)+\dd\delta(S_2)=\delta(\ell_1+\partial S_2)$. Figure~\ref{fig:twist} shows an example of such a deformation
of~$z$ on~$\partial M_4$, where the red square in the right panel represents
the surface $S_2$ corresponding to the deformation of~$z$.

In this way, the gauge transformation deforms the Wilson line supported on the
Poincar\'e dual of~$z$ through the coupling term $a\cup z+z\cup a$ in the
Chern--Simons action. The gauge transformation in~Eq.~\eqref{theta-stat}
captures the response to such a deformation, and gives the phase acquired by
the Wilson line when it undergoes a twisting deformation as depicted
in~Fig.~\ref{fig:twist}.\footnote{In other words, since the induced Wilson line
on the 't~Hooft loop winds around the hole, this deformation effectively counts
the number of local $2\pi$ rotations around the dyonic line operator.} Since
$r\cup z+z\cup r-r\cup\dd r$ takes integer values, this implies that the Wilson
line exhibits anyonic statistics with topological
spin~$\theta/(2\cdot2\pi)$.\footnote{This analysis corresponds to the
discussion on anyonic statistics in 3D lattice Chern--Simons theory
in~Ref.~\cite{Jacobson:2023cmr}.}\footnote{However, for general values of the
$\theta$-angle, due to the presence of the bulk term $(\theta/2)\sum_hz\cup z$,
each Wilson line becomes a non-genuine line operator, and therefore does not
imply a true any``on''.}

\subsection{Genuine loop operator on the lattice}
\label{genuine}
Let us consider the case where $m\theta=2\pi k$ with~$k\in\mathbb{Z}$. In this
case, in the continuum theory, the Wilson line induced by the Witten effect is
given by~$im\theta/(2\pi)\int_\mathcal{R}f=ik\int_\mathcal{R}f=ik\oint_{\partial\mathcal{R}}a\bmod2\pi$, and thus it should be a \emph{genuine loop
operator}~\cite{Kapustin:2014gua}. However, the operator~$w_i$ defined
in~Eq.~\eqref{unnit-Wilson-line} generally involves an associated
surface~$\mathcal{R}_i$ (and it is a component of the
$\theta$~term~\eqref{CS-w/-bulk-rewrite-R}). At finite lattice spacings, each
$w_i$ (and~$w_i'$) can fluctuate independently
in~Eq.~\eqref{CS-w/-bulk-rewrite-R}\footnote{In addition, the terms:
\begin{align}
   \frac{i\theta}{4\pi}\sum_{c\in\partial M_4}\sum_iq_i
   \left\{a\cup\delta(\partial\mathcal{R}_i)+\delta(\partial\mathcal{R}_i)
   \cup a\right\}_c+\frac{i\theta}{4\pi}\sum_{h\in M_4}
   \sum_iq_i\left\{2\pi z\cup\delta(\mathcal{R}_i)+2\pi\delta(\mathcal{R}_i)
   \cup z\right\}_h
\end{align}
in Eq.~\eqref{CS-w/-bulk-rewrite-R} can be rewritten as 
\begin{align}
   \sum_iq_i (w_i+w'_i).
\end{align}
In general, $w_i=w_j$ (and $w_i=w'_j$, $w'_i=w'_j$) does not holds at finite
lattice spacings. (If it holds, we can justify the calculation similar to the
continuum theory
like~$im\theta/(2\pi)\int_\mathcal{R}f=ik\int_\mathcal{R}f=ik\oint_{\partial\mathcal{R}}a\bmod2\pi$.)}, so it is not obvious whether this surface dependence cancels
out unlike the continuum theory.

Here, we show that the dyonic loop operator defined by the excision method
becomes a genuine loop operator already on the lattice without taking the
continuum limit. First, we note that whether the operator is genuine or not is
determined by the property of the last term of~Eq.~\eqref{CS}, because the
first term of the right-hand side consists of variables defined on the
hypersurface~$\partial M_4$ along the loop and thus the contribution of that
term is genuine. Then, in the last term of~Eq.~\eqref{CS}, we substitute $z$ by
the decomposition~\eqref{z-separate}, $z=z'+z_0$, where
\begin{align}
   z'\equiv\sum_iq_i\delta(\mathcal{R}_i).
\label{definition zprime}
\end{align}
$z'$ is the part of~$z$ which contains the information how the definition
of the dyonic operator depends on the surface~$\mathcal{R}_i$ spanned by the
Wilson loop~$w_i$. We note that the admissibility implies~$\dd z=0$. Since
$\dd z_0=0$ (recall that $z_0$ is given by the delta function whose support is
a union of closed surfaces), $z'$ is also flat,
\begin{align}
   \dd z'=0.
\label{flatness zprime}
\end{align}
We also note that, among the components of $z$, the part $z'$ contributes to the magnetic charge of the 't~Hooft loop, while $z_0$ does not. That is, the following relations hold:
\begin{align}
&\sum_{p\in\partial\mathcal{B}} z'_p= m, \label{magnetic-charge-contribution-zprime}\\
&\sum_{p\in\partial\mathcal{B}} (z_0)_p= 0. \label{magnetic-charge-contribution-z_0}
\end{align}
Here, $\mathcal{B}$ denotes the hole representing the monopole, as mentioned in Sect.~\eqref{excision-method}.

Here, by substituting the decomposition $z = z' + z_0$ into $z \cup z$ in Eq.~\eqref{CS}, we obtain
\begin{align}
    z\cup z&=z'\cup z'+z'\cup z_0+z_0\cup z'+z_0\cup z_0\notag\\
    &(=z'\cup z'+ 2z'\cup z_0- \dd(z'\cup_1 z_0)+z_0\cup z_0).
\label{decomposition-zcupz}
\end{align}
 Here, we used the flatness of $z'$~\eqref{flatness zprime} and a property of cup product~\eqref{cup-product-noncommutative}.
 We now show that the
term~$\sum_h(z'\cup z'+z'\cup z_0+z_0\cup z')_h$ which potentially depends on the
surfaces~$\mathcal{R}_i$ through~Eq.~\eqref{definition zprime} can actually be
expressed solely by degrees of freedom on the boundary~$\partial M_4$; this
implies that the loop operator is genuine.

\subsubsection{Case of even $m$}
\label{even-m}
Let us consider the case where $m\theta=2\pi k$, and $m$ is even. We show that,
in this case,
$e^{\frac{i\theta}{2}\sum_h(z'\cup z'+z'\cup z_0+z_0\cup z')_h}=e^{\frac{ik\pi}{m}\sum_h(z'\cup z'+z'\cup z_0+z_0\cup z')_h}$ can be
rewritten by using solely degrees of freedom defined on the
boundary~$\partial M_4$. As a preparation, we define a Chern--Simons-like
partition function on a closed 3D manifold~$\mathcal{M}_3$ as follows:
\begin{align}
\mathbf{Z}_{\mathcal{M}_3}^k[z',z_0]&\equiv
e^{\frac{ik\pi}{N}\sum_{c\in \mathcal{M}_3}(z_0\cup_1z')_c}\notag\\
&\qquad{}\times
\left(\prod_{\ell\in\mathcal{M}_3}\frac{1}{m}\sum_{c_\ell=0}^{m-1}\right)
   \delta_m[(\dd c_\ell-z')_p]e^{-\frac{ik\pi}{m}\sum_{c\in\mathcal{M}_3}
   (-c\cup\dd c+z'\cup c+c\cup z'+2c\cup z_0)_c}.
\label{m-even-noninv-defect}
\end{align}
Through discussion of this section, let the 3D manifold $\mathcal{M}_3$ represent the 't~Hooft loop, e.g., $\mathcal{M}_3\cong\partial M_4\cong S^2\times S^1$. Here, the delta function~$\delta_m[(\dd c_\ell-z')_p]$ is explicitly given by a
Kronecker delta~$\prod_{p\in\mathcal{M}_3}\delta_{(\dd c_\ell-z')_p,m\mathbb{Z}}$. When
$\mathbf{Z}_{\mathcal{M}_3}^k[z']$ is non-vanishing, there exist integer-valued
fields $\mu$ and~$\nu$ such that
\begin{align}
   z'=\dd\mu+m\nu
\label{zprime trivialized}
\end{align}
without loss of generality. This is consistent with the boundary condition on $\partial\mathcal{B}$~\eqref{def:magnetic-charge}. We note that the integer valued field~$\nu$ is flat
from~Eq.\eqref{flatness zprime}. Therefore, we can rewrite the partition
function as
\begin{align}
   &\mathbf{Z}_{\mathcal{M}_3}^k[z',z_0]
\notag\\
   &=e^{-\frac{ik\pi}{m}\sum_{c\in\mathcal{M}_3}(\mu\cup\dd\mu+m\mu\cup\nu+m\nu\cup\mu+2\mu\cup z_0-z_0\cup_1z')_c}\notag\\
&\qquad{}\times\left(\prod_{\ell\in\mathcal{M}_3}\frac{1}{m}\sum_{c_\ell=0}^{m-1}\dd c_\ell\right)
   \delta_m[(\dd c_\ell)_p]\,e^{-\frac{ik\pi}{m}\sum_{c\in\mathcal{M}_3}(-c\cup\dd c+2c\cup z_0)_c}
\notag\\
   &=e^{-\frac{ik\pi}{m}\sum_{c\in\mathcal{M}_3}(\mu\cup\dd\mu+m\mu\cup\nu+m\nu\cup\mu+2\mu\cup z_0-z_0\cup_1z')_c}\mathbf{Z}_{\mathcal{M}_3}^k[z'=0, z_0],
\label{m-even-noninv-defect zprime trivialized}
\end{align}
where, in the first equality, we have made the shift, $c\to c+\mu$ by using
$\mu$ in~Eq.~\eqref{zprime trivialized}. In the exponent, we have dropped the  term~$mc\cup\nu+m\nu\cup c$, because, under the
constraint~$\delta_m[(\dd c_\ell)_p]\to\dd c=0\bmod m$, the relation~$\dd c=m\dd g$ holds for an integer-valued field~$g$; thus
$mc\cup\nu+m\nu\cup c=\dd(\nu\cup_1 c)+2mc\cup\nu-m^2\nu\cup\dd g$ becomes a total derivative up to~$2m\mathbb{Z}$
(Such terms do not contribute to Boltzmann factor). Substituting
$\dd c=m\dd g$ into $c\cup\dd c$ shows that
$c\cup\dd c=-m\dd(c\cup g)+m^2\dd g\cup g$. This implies that the
factor in the third line
of~Eq.~\eqref{m-even-noninv-defect zprime trivialized} is unity,
$e^{-\frac{ik\pi}{m}\sum_{c\in\mathcal{M}_3}(-c\cup\dd c)_c}=1$. Finally, let us evaluate the remaining factor $e^{-\frac{2k\pi i}{m}\sum_{c\in\mathcal{M}_3}(c\cup z_0)_c}$. First, we consider the identity~$e^{-\frac{2k\pi i}{m}\sum_{c\in\mathcal{M}_3}(c\cup z_0)_c} = e^{-\frac{2k\pi i}{m}\sum_{c\in\mathcal{M}_3}((c - mg)\cup z_0)_c}.$
Furthermore, we find~$\sum_{c\in\mathcal{M}_3}((c - mg)\cup z_0)_c = \sum_{c\in\text{Poincar\'e dual of $c - mg$}} z_0.$
Here, since $\dd c = m \dd g$, we have $\dd(c - mg) = 0$, and therefore the Poincar\'e dual of $c - mg$ is a closed surface on the 3D closed manifold~$\mathcal{M}_3$. Noting that closed surfaces on $\mathcal{M}_3$ can be either contractible or non-contractible, the sum~$\sum_{c\in\text{Poincar\'e dual of $c - mg$}} z_0$ can be decomposed as
\begin{align}
\sum_{c\in\text{Poincar\'e dual of $c - mg$}} z_0 = \sum_{c\in\text{contractible}} z_0 + \sum_{c\in\text{non-contractible}} z_0.
\end{align}
Here, the first term on the right hand side vanishes due to $\dd z_0 = 0$. Among closed surfaces on $\mathcal{M}_3\cong S^2\times S^1$, the only non-contractible ones are the boundaries of holes~(monopole), denoted by $\partial\mathcal{B}\cong S^2$. Therefore, from Eq.~\eqref{magnetic-charge-contribution-z_0}, the second term also vanishes. Thus, we conclude that~$
e^{-\frac{2k\pi i}{m}\sum_{c\in\mathcal{M}_3}(c\cup z_0)_c} = 1$, and $\mathbf{Z}_{\mathcal{M}_3}^k[z'=0, z_0]=\left(\prod_{\ell\in\mathcal{M}_3}\frac{1}{m}\sum_{c_\ell=0}^{m-1}\dd c_\ell\right)
   \delta_m[(\dd c_\ell)_p].$

Then, we can invoke the
computation of the partition function in~Ref.~\cite{Honda:2024sdz} and
infer that
\begin{align}
\mathbf{Z}_{\mathcal{M}_3}^k[z'=0, z_0]&=\left(\prod_{\ell\in\mathcal{M}_3}\frac{1}{m}\sum_{c_\ell=0}^{m-1}\dd c_\ell\right)
   \delta_m[(\dd c_\ell)_p]\notag\\
   &=m^{b_2-1}\label{Partition-function-Betti}
\end{align}
where $b_2$~denotes the
second Betti number of~$\mathcal{M}_3$. Since $\mathbf{Z}_{\mathcal{M}_3}^k[0]$ is
given by the Betti number, it is invariant under a topological deformation
$\mathcal{M}_3\to\mathcal{M}_3'$. Consequently, we obtain the following
relation:
\begin{align}
   \mathbf{Z}_{\mathcal{M}_3'}^k[z',z_0]
   &=e^{-\frac{ik\pi}{m}\sum_{c\in\mathcal{M}_3'\cup\overline{\mathcal{M}_3}}
(\mu\cup\dd\mu+m\mu\cup\nu+m\nu\cup\mu+2\mu\cup z_0-z_0\cup_1z')_c}\notag\\
&\qquad\times e^{-\frac{ik\pi}{m}\sum_{c\in\mathcal{M}_3}(\mu\cup\dd\mu+m\mu\cup\nu+m\nu\cup\mu+2\mu\cup z_0-z_0\cup_1z')_c}
   \mathbf{Z}_{\mathcal{M}_3}^k[z'=0,z_0]
\notag\\
   &=e^{-\frac{ik\pi}{m}\sum_{c\in\mathcal{M}_3'\cup\overline{\mathcal{M}_3}}
   (\mu\cup\dd\mu+m\mu\cup\nu+m\nu\cup\mu+2\mu\cup z_0-z_0\cup_1z')_c}
   \mathbf{Z}_{\mathcal{M}_3}^k[z',z_0]
\notag\\
   &=e^{-\frac{ik\pi}{m}\sum_{h\in\mathcal{V}_4}
   \dd(\mu\cup\dd\mu+m\mu\cup\nu+m\nu\cup\mu+2\mu\cup z_0-z_0\cup_1z')_c}
   \mathbf{Z}_{\mathcal{M}_3}^k[z',z_0]
\notag\\
   &=e^{-\frac{ik\pi}{m}\sum_{h\in\mathcal{V}_4}(z'\cup z'+z'\cup z_0+z_0\cup z')_h}
   \mathbf{Z}_{\mathcal{M}_3}^k[z',z_0].
\label{m-even-top-deform}
\end{align}
Here, $\mathcal{V}_4$ denotes a bulk manifold satisfying
$\partial\mathcal{V}_4=\mathcal{M}_3'\cup\overline{\mathcal{M}_3}$, and we used the topological invariance~$\mathbf{Z}_{\mathcal{M}_3}^k[0]=\mathbf{Z}_{\mathcal{M}_3'}^k[0].$ Since the
't~Hooft loop is defined on the boundary of~$M_4$, we regard $\mathcal{M}_3$ as
a 't~Hooft loop, $\mathcal{M}_3=\partial M_4$. We now consider the
deformation~$\mathcal{M}_3\to\mathcal{M}'_3$ based
on~Eq.~\eqref{m-even-top-deform}.

For simplicity, consider two 't~Hooft loops (a monopole–anti-monopole pair)
along the time direction. Let $M_3^1$ and~$M_3^2$ be hypersurfaces enclosing each of the 't~Hooft loops respectively,
and define $\mathcal{M}_3'=M_3^1\cup M_3^2$. At first, we assume that $M_3^1$ and $M_3^1$ have no intersection~$M_3^1\cap M_3^2=\varnothing$. Then the partition function~$\mathbf{Z}_{\mathcal{M}_3'}^k[z',z_0]$ can be represented as
\begin{align}
\mathbf{Z}_{\mathcal{M}_3'}^k[z',z_0]=\mathbf{Z}_{M^1_3}^k[z',z_0]\mathbf{Z}_{M^2_3}^k[z',z_0].\label{decomposition-partition-function}
\end{align}

\begin{figure}[htbp]
\centering
\includegraphics[scale=0.59]{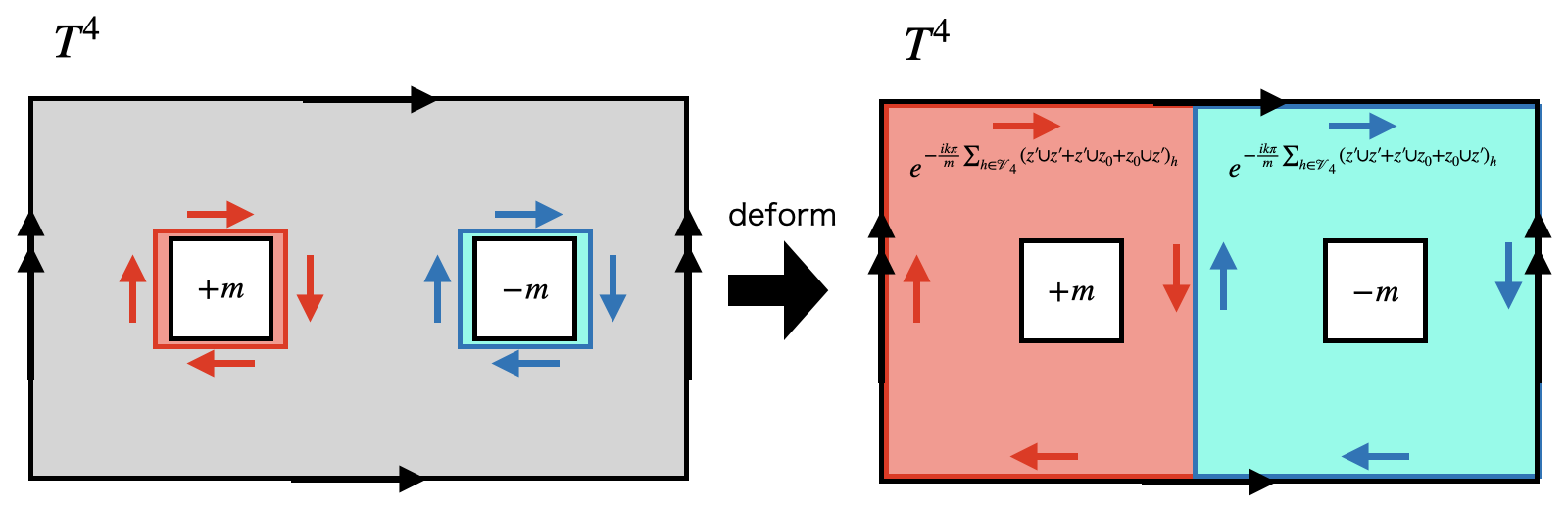}
\caption{Deformation of $M^1_3$ and $M^2_3$ viewed from a 2D slice. The largest square represents the lattice manifold $M_4 \cong T^4$, with arrows on its edges indicating identifications corresponding to the periodic boundary conditions of the torus. The regions labeled $+m$ and $-m$ at the center denote monopoles with magnetic charges $+m$ and $-m$, respectively, defined via the excision method. The red and blue squares surrounding the monopoles represent $M^1_3$ and $M^2_3$, respectively, and the colored arrows around each square indicate their orientations. The pale-colored regions within the red and blue squares represent the deformation of $M^1_3$ and $M^2_3$ according to Eq.~\eqref{m-even-top-deform}, producing the factor $e^{-\frac{ik\pi}{m}\sum_{h\in\mathcal{V}_4}(z'\cup z'+z'\cup z_0+z_0\cup z')_h}$. From the left to the right figure, the red and blue squares are deformed such that the pale-colored region $\mathcal{V}_4$ eventually covers the entire manifold, i.e., $\mathcal{V}_4 = M_4$.
In this case, taking into account the periodic boundary conditions of the torus and focusing on the red and blue squares corresponding to $M^1_3$ and $M^2_3$, respectively, one finds that the faces with opposite orientations are facing each other.
}
\label{fig:deform}
\end{figure}

Let us consider expanding $M^1_3$ and $M^2_3$ until $\mathcal{V}_4 = M_4$ is achieved. Using Eqs.~\eqref{m-even-noninv-defect zprime trivialized} and \eqref{decomposition-partition-function}, we obtain
\begin{align}
\mathbf{Z}_{\mathcal{M}_3'}^k[z',z_0] &= \mathbf{Z}_{M^1_3}^k[z',z_0] \mathbf{Z}_{M^2_3}^k[z',z_0] \notag \\
&= e^{-\frac{ik\pi}{m} \sum_{c \in M^1_3} (\mu \cup \dd\mu + m\mu \cup \nu + m\nu \cup \mu + 2\mu \cup z_0 - z_0 \cup_1 z')_c} \notag \\
&\quad \times e^{-\frac{ik\pi}{m} \sum_{c \in M^2_3} (\mu \cup \dd\mu + m\mu \cup \nu + m\nu \cup \mu + 2\mu \cup z_0 - z_0 \cup_1 z')_c} \notag \\
&\quad \times \mathbf{Z}_{M^1_3}^k[z'=0, z_0] \mathbf{Z}_{M^2_3}^k[z'=0, z_0]
\end{align}

Then, as illustrated in Fig.~\eqref{fig:deform}, cancellations occur. Specifically, due to the disjoint support $M_3^1 \cap M_3^2 = \varnothing$ and the self-overlap of $M^1_3$ and $M^2_3$ arising from the periodicity of the torus, the factors~$e^{-\frac{ik\pi}{m} \sum_{c \in M^1_3} (\mu \cup \dd\mu + m\mu \cup \nu + m\nu \cup \mu + 2\mu \cup z_0 - z_0 \cup_1 z')_c}$ and $e^{-\frac{ik\pi}{m} \sum_{c \in M^2_3} (\mu \cup \dd\mu + m\mu \cup \nu + m\nu \cup \mu + 2\mu \cup z_0 - z_0 \cup_1 z')_c}$ mutually cancel (see Fig.~\eqref{fig:deform}). Therefore, we obtain
\begin{align}
\mathbf{Z}_{\mathcal{M}_3'}^k[z',z_0] = \mathbf{Z}_{M^1_3}^k[z'=0,z_0] \mathbf{Z}_{M^2_3}^k[z'=0,z_0] \label{after^cancelation}
\end{align}

Since the topology of the 't~Hooft loop defined by the excision method is $S^2 \times S^1$, both $M^1_3$ and $M^2_3$ are topologically equivalent to $S^2 \times S^1$. Thus, by using Eq.~\eqref{Partition-function-Betti}, we find
\begin{align}
\mathbf{Z}_{\mathcal{M}_3'}^k[z',z_0]
&= \mathbf{Z}_{M^1_3}^k[z'=0,z_0] \mathbf{Z}_{M^2_3}^k[z'=0,z_0] \notag \\
&= \mathbf{Z}_{S^2 \times S^1}^k[z'=0,z_0] \mathbf{Z}_{S^2 \times S^1}^k[z'=0,z_0] = 1
\end{align}

Therefore, referring to Eq.~\eqref{m-even-top-deform}, we conclude that when $M^1_3$ and $M^2_3$ are expanded until $\mathcal{V}_4 = M_4$,
\begin{align}
\mathbf{Z}_{\mathcal{M}_3'}^k[z',z_0]&= 1 \notag \\
&=e^{-\frac{ik\pi}{m} \sum_{h \in \mathcal{V}_4}(z' \cup z' + z' \cup z_0 + z_0 \cup z')_h} \, \mathbf{Z}_{\mathcal{M}_3}^k[z',z_0].
\end{align}

Now, we finally arrive at the conclusion,
\begin{align}
   e^{\frac{ik\pi}{m}\sum_{h\in M_4}(z'\cup z'+z'\cup z_0+z_0\cup z')_h}
=\mathbf{Z}_{\mathcal{M}_3=\partial M_4(=\text{'t Hooft loop})}^k[z', z_0].
\end{align}
In other words, when 't~Hooft loop~$\partial M_4$ exists, by using the
decomposition~\eqref{decomposition-zcupz}, the
$\theta(=2\pi k/m)$~term~\eqref{CS} can be rewritten as
\begin{align}
   &\exp{\frac{2\pi ki}{4\pi m}\sum_{c\in\partial M_4(=\text{'t Hooft loop})}
   \left(\frac{a\cup\dd a}{2\pi}+a\cup z+z\cup a\right)_c
   +\frac{ik\pi}{m}\sum_{h\in M_4}(z\cup z)_h}
\notag\\
   &=\exp{\frac{2\pi ki}{4\pi m}\sum_{c\in\partial M_4(=\text{'t Hooft loop})}
   \left(\frac{a\cup\dd a}{2\pi}+a\cup z+z\cup a\right)_c}
   \mathbf{Z}_{\mathcal{M}_3=\partial M_4(=\text{'t Hooft loop})}^k[z', z_0]
\notag\\
   &\qquad{}
   \times e^{\frac{ik\pi}{m}\sum_{h\in M_4}(z_0\cup z_0)_h}.
\label{rewrite-CS-boundary}
\end{align}
The left-hand side, via~Eqs.~\eqref{CS} and~\eqref{CS-w/-bulk-rewrite-R},
depends on the non-genuine Wilson loops given
in~Eqs.~\eqref{unnit-Wilson-line} and~\eqref{unnit-Wilson-line2}, whereas the
right-hand side does not; the dependence on~$z'$ as defined
in~Eq.~\eqref{definition zprime} is expressed solely in terms of quantities
defined along the 't~Hooft loop~$\partial M_4$. In other words, on the
right-hand side, the non-genuine Wilson loops
in~Eqs.~\eqref{unnit-Wilson-line} and~\eqref{unnit-Wilson-line2} have been
rewritten as genuine loop operators. Therefore,
Eq.~\eqref{rewrite-CS-boundary} demonstrates that the dyonic operator arising
from Wilson loops induced via the Witten effect becomes a genuine loop operator
when $m\theta=2\pi k$ and $m$~is even, even at finite lattice spacings.

\subsubsection{Case of odd $m$}
Next, we consider the case where $m\theta=2\pi k$, and $m$ is odd. We show
that, in this case also, $e^{\frac{i\theta}{2}\sum_h(z'\cup z'+z'\cup z_0+z_0\cup z')_h}=e^{\frac{ik\pi}{m}\sum_h(z'\cup z'+z'\cup z_0+z_0\cup z')_h}$ can be rewritten by using solely degrees of freedom defined on the
boundary~$\partial M_4$. Here, we consider the following decomposition:
\begin{align}
   &\exp{\frac{i\pi k}m\sum_{h\in M_4}(z'\cup z'+z'\cup z_0+z_0\cup z')_h}
\notag\\
   &=\exp{\frac{i\pi k(2m+1)}m\sum_{h\in M_4}(z'\cup z'+z'\cup z_0+z_0\cup z')_h}
\notag\\
   &=\exp{i\pi k\sum_{h\in M_4}(z'\cup z')_h}\times
   \exp{\frac{i\pi k(m+1)}m\sum_{h\in M_4}(z'\cup z'+2z'\cup z_0)_h}\notag\\
   &\qquad{}\times\exp{-\frac{i\pi k}m\sum_{h\in M_4}\dd(z'\cup_1z_0)_c}
\label{decomposition}
\end{align}
Now, regarding the term $\exp{i\pi k\sum_{h\in M_4}(z'\cup z')_h}$, it has been
shown in~Ref.~\cite{Chen:2019mjw,Gaiotto:2015zta} that when $z'$ is
flat~(i.e.~Eq.~\eqref{flatness zprime}),
\begin{align}
   &\exp{i\pi\sum_{h\in M_4}(z'\cup z')_h}
\notag\\
   &=e^{i\sum_{p\in\Sigma}z'_p}\left(\prod_{p\in\partial M_4}
   \int\dd\gamma_p\,\dd\bar{\gamma}_p\right)
   \left(\prod_{c\in\partial M_4}h_c[z']\right)
   \left(\prod_{p\in\partial M_4}\left(1+\bar\gamma_p\gamma_p\right)\right)
\notag\\
   &\equiv e^{i\sum_{p\in\Sigma}z'_p}\mathcal{Z}_{\gamma,\partial M_4(=\text{'t~Hooft loop})}[z'],
\label{Z_gamma_def}
\end{align}
by using the partition function corresponding to the world line of a Majorana
fermion along the Poincar\'e dual of~$z'$, $e^{i\sum_{p\in\Sigma}z'_p}\mathcal{Z}_{\gamma,\partial M_4(=\text{'t~Hooft loop})}[z']$.
Here, $\gamma_p$ and~$\bar{\gamma_p}$ are Grassmann number integration
variables defined on plaquettes. For the precise definition of~$h_c[z']$, see
Ref.~\cite{Chen:2019mjw}. The factor~$e^{i\sum_{p\in\Sigma}z'_p}$ counts the
intersection between a two-dimensional surface~$\Sigma$ and the
Poincar\'e dual of~$z'$ on~$\partial M_4$; this controls whether the boundary
condition for the fermion~$\gamma_p$ along the worldline~$S^1$ representing the
't~Hooft loop is periodic or anti-periodic~\cite{Gaiotto:2015zta,Chen:2019mjw}.
That is, it represents the background data specifying the spin structure on the
't~Hooft loop. The rewriting~\eqref{Z_gamma_def} reflects the fermionic
statistics of the Wilson line induced on the 't~Hooft loop.

Furthermore, from the discussion in the previous
Sect.~\eqref{even-m}, the remaining factor
in~Eq.~\eqref{decomposition}, for $m+1$ even, can be expressed based on the partition
function of the level-$m$ $\mathbb{Z}_m$ BF theory defined
in~Refs.~\cite{Honda:2024sdz,Honda:2024yte} as
\begin{align}
&\exp{\frac{i\pi k(m+1)}m\sum_{h\in M_4}(z'\cup z'+2z'\cup z_0)_h}\times\exp{-\frac{i\pi k}m\sum_{h\in M_4}\dd(z'\cup_1z_0)_c}\notag\\
&\qquad=\mathcal{Z}^{k(m+1)}_{\partial M_4(=\text{'t Hooft loop})}[z', z_0],
\label{BF-rewrite}
\end{align}
where, explicitly\footnote{%
A lattice counterpart of the topological defect for the non-invertible axial
symmetry~\cite{Choi:2022jqy,Cordova:2022ieu} can be constructed by using
this partition function~\cite{Honda:2024sdz,Honda:2024yte}.}
\begin{align}
\mathcal{Z}^{k(m+1)}_{\partial M_4(=\text{'t Hooft loop})}[z', z_0]&\equiv \exp{\frac{i\pi k}m\sum_{h\in\partial M_4}(z'\cup_1z_0)_c}\notag\\
&\times\left(\prod_{\ell\in \partial M_4}\frac{1}m\sum_{c_\ell=0}^{N-1}\right)
   \delta_N[(\dd c_\ell-z')_p]e^{-\frac{ik(m+1)\pi}m\sum_{c\in \partial M_4}(z'\cup c+2c\cup z_0)_c}.
\label{BF-def}
\end{align}
This partition function~$\mathcal{Z}^{k(m+1)}_{\partial M_4(=\text{'t Hooft loop})}[z', z_0]$ preserves the
property~\eqref{m-even-top-deform}. By following exactly the same line of
reasoning in the previous Sect.~\eqref{even-m}, one can
derive~Eq.~\eqref{BF-rewrite}.

To summarize, we have shown
\begin{align}
   &\exp{\frac{i\pi k}m\sum_{h\in M_4}(z'\cup z'+z'\cup z_0+z_0\cup z')_h}
\notag\\
   &=\exp{i\pi k\sum_{h\in M_4}(z'\cup z')_h}\times
   \exp{\frac{i\pi k(m+1)}m\sum_{h\in M_4}(z'\cup z'+2z'\cup z_0)_h}\notag\\
   &\qquad{}\times\exp{-\frac{i\pi k}m\sum_{h\in M_4}\dd(z'\cup_1z_0)_c}\notag\\
   &=(e^{i\sum_{p\in\Sigma}z'_p}\mathcal{Z}_{\gamma,\partial M_4(=\text{'t~Hooft loop})}[z'])^k\times
\mathcal{Z}^{k(m+1)}_{\partial M_4(=\text{'t Hooft loop})}[z', z_0].
\label{N-odd-bulk-rewrite}
\end{align}
Therefore, even in the case of odd~$m$, the factor~$\exp{\frac{i\pi k}m\sum_{h\in M_4}(z'\cup z'+z'\cup z_0+z_0\cup z')_h}$ can be
expressed entirely in terms of degrees of freedom defined only on the
boundary. In other words, corresponding to~Eq.~\eqref{rewrite-CS-boundary}, we
have shown
\begin{align}
   &\exp{\frac{2\pi ki}{4\pi m}\sum_{c\in\partial M_4(=\text{'t Hooft loop})}
   \left(\frac{a\cup\dd a}{2\pi}+a\cup z+z\cup a\right)_c
   +\frac{ik\pi}{m}\sum_{h\in M_4}(z\cup z)_h}
\notag\\
   &=\exp{\frac{2\pi ki}{4\pi m}\sum_{c\in\partial M_4(=\text{'t Hooft loop})}
   \left(\frac{a\cup\dd a}{2\pi}+a\cup z+z\cup a\right)_c}
\notag\\
   &\qquad{}
   \times(e^{i\sum_{p\in\Sigma}z'_p}\mathcal{Z}_{\gamma,\partial M_4(=\text{'t~Hooft loop})}[z'])^k\times
\mathcal{Z}^{k(m+1)}_{\partial M_4(=\text{'t Hooft loop})}[z', z_0]
   \times e^{\frac{ik\pi}{m}\sum_{h\in M_4}(z_0\cup z_0)_h}
\label{rewrite-CS-boundary-m-odd}.
\end{align}
Similar to the discussion in~Sect.~\ref{even-m}, the rewriting
in~Eq.~\eqref{rewrite-CS-boundary-m-odd} demonstrates that the dyonic
operator arising from Wilson loops induced via the Witten effect becomes a
genuine loop operator when $m\theta=2\pi k$ and $m$ is odd. 

\section{Conclusion}
\label{conclusion}
In this paper, we introduced the 't~Hooft loop in 4D Maxwell theory using an
excision method that respects the admissibility condition. In this formulation,
the statistics of dyons induced by the Witten effect can be readily extracted
as the gauge transformation property of the boundary Chern--Simons term
resulting from the $\theta$ term. In the fully regularized lattice framework,
the Wilson line appears directly as a coefficient of the gauge field~$z$ in the
Chern--Simons term.

Moreover, since the 't~Hooft line is defined as a hole with finite size, the
Wilson line can twist on the surface of the hole under gauge transformations.
From the gauge variation of the Chern--Simons term, the topological spin
encodes the statistical property arising from the twist of the Wilson line.
We have further demonstrated that, within this lattice formulation, when
when $m\theta=2\pi\mathbb{Z}$, the dyonic loop becomes a genuine loop operator
even at finite lattice spacings.

Moreover, although it is not directly related to the main topic of this paper, we mention that the identities shown in subsection~\eqref{genuine}, namely Eqs.~\eqref{rewrite-CS-boundary} and \eqref{rewrite-CS-boundary-m-odd} can be applied to the lattice formulation of non-invertible axial symmetry defects. In Ref.~\cite{Honda:2024yte}, only the identity corresponding to the case $k \in 2\mathbb{Z}$ in Eq.~\eqref{rewrite-CS-boundary-m-odd} was derived. Consequently, the lattice formulation of non-invertible axial symmetry defects was realized only for the case where the axial rotation angle $2\pi p/N$ satisfies $p \in 2\mathbb{Z}$ with $\gcd(p,N) = 1$. However, the identities \eqref{rewrite-CS-boundary} and \eqref{rewrite-CS-boundary-m-odd} demonstrate that the construction in Ref.~\cite{Honda:2024yte} can be generalized to arbitrary rational axial rotation angles $2\pi p/N$.

Let us briefly mention an alternative approach based on the modified Villain
formulation that does not directly impose the admissibility condition
(e.g., Refs.~\cite{Jacobson:2023cmr,Anosova:2022cjm,Honda:2024sdz}). In this
approach, the Bianchi identity is imposed via a dual gauge field and the
't~Hooft line can be naturally defined as its line operator. The Witten effect
can then be naturally derived from the requirement of gauge
invariance~\cite{Jacobson:2023cmr,Honda:2024sdz}. The spin of line operators
could also be identified as the response to a change in the definition of the
cup product~\cite{Xu:2024hyo}.

We now comment on possible extensions of the results presented in this work. In
this paper, we have focused on the case where the charge-$1$ Wilson loop and
the 't~Hooft loop are bosonic. More generally, situations in which these
operators are fermionic may also be considered. In such cases, it is known that
the spin of the dyon is also shifted by~$+1/2$. To analyze these cases, it is
necessary to explicitly define the second Stiefel--Whitney class on the
lattice.\footnote{For simplicial lattices, see Ref.~\cite{GT76}.} In
particular, it is of interest to study the case of non-spin manifolds where
the second Stiefel--Whitney class is nontrivial (e.g.,
Ref.~\cite{Kan:2024fuu}).

Even for non-spin manifolds, it is expected that the 't~Hooft line can be
defined via the excision method, at least for simplicial lattices; we hope to
explore this direction in future work.

It is also interesting to consider extending our discussion to non-abelian
gauge theories such as~$PSU(N)$~\cite{Abe:2023ncy}. The excision method that
respects the admissibility condition seems compatible with L\"uscher’s
construction of the principal bundle~\cite{Luscher:1981zq}. If the
Chern--Simons term can be extracted from the local expression of the
topological charge~\cite{Luscher:1981zq}, a study parallel to the present work
may be carried out; see a related work in~Ref.~\cite{Zhang:2024cjb}.

The author hopes that, in the future, the admissibility-respecting formulation
considered in the present work could be generalized to magnetic objects in
lattice formulation of non-abelian gauge
theory~\cite{Luscher:1981zq,Abe:2023ncy} and chiral gauge
theory~\cite{Luscher:1998du}.

\section*{Acknowledgments}
The author would like to thank Motokazu Abe, Yamato Honda, Okuto Morikawa,
Hiroshi Suzuki, and Yuya Tanizaki for useful discussions and collaborations.
The author also would like to thank Hiroki Wada and Ryo Yokokura for useful
discussions. The author would like to especially thank Hiroshi Suzuki for
a careful reading of the manuscript and providing me insightful comments.
This work was partially supported by Kyushu University’s Innovator Fellowship
Program and by the JSPS KAKENHI, Grant Number~JP25KJ1954.

\appendix
\section{Proof that $(1/2)\sum_{h\in M_4}(z\cup z)_h\in\mathbb{Z}$}
\label{integer}

We prove that $(1/2)\sum_{h\in M_4}(z\cup z)_h\in\mathbb{Z}$ when $M_4\cong T^4$. Under the admissibility condition, the gauge field configuration satisfies $\dd z = 0$, so the Poincar\'e dual of~$z$ is a collection of closed two-dimensional surfaces. Among these, the contributions that cannot be gauged away are topologically two-dimensional tori~$T^2$ that nontrivially wrap around the cycles of~$M_4\cong T^4$.

In general, such a configuration can be expressed using a collection of closed two-dimensional surfaces labeled by indices $i$, $j$, $k$, $\ell$, $m$, and $n$: $\mathcal{M}^{1}_i\cong (T^2)_{12}$, $\mathcal{M}^{2}_j\cong (T^2)_{23}$, $\mathcal{M}^{3}_k\cong (T^2)_{34}$, $\mathcal{M}^{4}_\ell\cong (T^2)_{14}$, $\mathcal{M}^{5}_m\cong (T^2)_{13}$, and $\mathcal{M}^{6}_n\cong (T^2)_{24}$,\footnote{$(T^2)_{ab}$ denotes the two-dimensional torus spanned by the unit basis vectors $\hat{a}$ and $\hat{b}$ of $M_4\cong T^4$.} so that
\begin{align}
    z&=\sum_ip_i^1\delta(\mathcal{M}^{1}_i)+\sum_jp_j^2\delta(\mathcal{M}^{2}_j)+\sum_kp_k^3\delta(\mathcal{M}^{3}_k)+\sum_\ell p_\ell^4\delta(\mathcal{M}^{4}_\ell)+\sum_mp_m^5\delta(\mathcal{M}^{5}_m)+\sum_np_n^6\delta(\mathcal{M}^{6}_n)\notag\\
    &\equiv\sum_{\alpha=1}^6\sum_{i_\alpha}p_{i_\alpha}^\alpha\delta(\mathcal{M}^{\alpha}_{i_\alpha}).
\end{align}
In the second line, we unified the indices $i$, $j$, $k$, $\ell$, $m$, and $n$ into a single index $i_\alpha$, with $\alpha$ labeling the type of the closed two-dimensional surfaces.\footnote{This expression is valid up to a gauge transformation. However, since $(1/2)\sum_{h\in M_4}(z\cup z)_h$ is gauge invariant when the manifold has no boundary, we are free to fix a particular gauge.} Here $p_{i_\alpha}^\alpha$ are integer valued coefficients.

Then we compute
\begin{align}
    z\cup z&=\sum_{\alpha}\sum_{\beta}\sum_{i_\alpha}\sum_{i_{\beta}}p_{i_\alpha}^\alpha p_{i_\beta}^\beta\delta(\mathcal{M}^{\alpha}_{i_\alpha})\cup\delta(\mathcal{M}^{\beta}_{i_\beta})\notag\\
    &=\sum_{\alpha\neq\beta}\sum_{i_\alpha}\sum_{i_{\beta}}p_{i_\alpha}^\alpha p_{i_\beta}^\beta\delta(\mathcal{M}^{\alpha}_{i_\alpha})\cup\delta(\mathcal{M}^{\beta}_{i_\beta})\notag\\
    &=\sum_{\alpha<\beta}\sum_{i_\alpha}\sum_{i_{\beta}}p_{i_\alpha}^\alpha p_{i_\beta}^\beta\left(\delta(\mathcal{M}^{\alpha}_{i_\alpha})\cup\delta(\mathcal{M}^{\beta}_{i_\beta})+\delta(\mathcal{M}^{\beta}_{i_\beta})\cup\delta(\mathcal{M}^{\alpha}_{i_\alpha})\right).
\end{align}
In going from the first to the second line, we used the fact that there are no self-intersections, so $\delta(\mathcal{M}^{\alpha}_{i_\alpha})\cup\delta(\mathcal{M}^{\alpha}_{i_\alpha}) = 0$. 
The third line follows from $\sum_{\alpha\neq\beta}=\sum_{\alpha<\beta}+\sum_{\alpha>\beta}$ and exchanging the dummy indices $\alpha \leftrightarrow \beta$.
Using Eq.~\eqref{cup-product-noncommutative}, one can show that
\begin{align}
    \delta(\mathcal{M}^{\alpha}_{i_\alpha})\cup\delta(\mathcal{M}^{\beta}_{i_\beta})+\delta(\mathcal{M}^{\beta}_{i_\beta})\cup\delta(\mathcal{M}^{\alpha}_{i_\alpha}) = 2\delta(\mathcal{M}^{\alpha}_{i_\alpha})\cup\delta(\mathcal{M}^{\beta}_{i_\beta}) - \dd(\delta(\mathcal{M}^{\beta}_{i_\beta})\cup_1\delta(\mathcal{M}^{\alpha}_{i_\alpha})).
\end{align}
Therefore, we obtain
\begin{align}
    \frac{1}{2}\sum_{h\in M_4}(z\cup z)_h=\sum_{h\in M_4}\sum_{\alpha<\beta}\sum_{i_\alpha}\sum_{i_{\beta}}p_{i_\alpha}^\alpha p_{i_\beta}^\beta\left\{\delta(\mathcal{M}^{\alpha}_{i_\alpha})\cup\delta(\mathcal{M}^{\beta}_{i_\beta})\right\}_h\in\mathbb{Z}.
\end{align}
Here, we used the fact that $M_4\cong T^4$ has no boundary, thus $\sum_{h\in M_4}\dd(\delta(\mathcal{M}^{\beta}_{i_\beta})\cup_1\delta(\mathcal{M}^{\alpha}_{i_\alpha}))=0$.

\bibliographystyle{ptephy}
\bibliography{ref}
\vspace{0.2cm}
\noindent

\let\doi\relax

\end{document}